\documentclass[a4paper,11pt]{article}
\usepackage{fullpage}
\usepackage{authblk}
\usepackage[urlcolor=blue,citecolor=black,linkcolor=black]{hyperref}
\usepackage{natbib}
\usepackage{graphicx}
\usepackage{amsmath}
\usepackage{amssymb}

\title{Relationship between the moment of inertia and the $k_2$ Love number of fluid extra-solar planets}

\author[1]{Anastasia Consorzi}
\author[2]{Daniele Melini}
\author[1]{Giorgio Spada}

\affil[1]{Dipartimento di Fisica e Astronomia ``Augusto Righi'' (DIFA), Alma Mater Studiorum Università di Bologna, Viale Berti Pichat, 8, 40127 Bologna, Italy}
\affil[2]{Istituto Nazionale di Geofisica e Vulcanologia, Via di Vigna Murata, 605, 00143 Roma, Italy}

\date{}

\begin{document}

\maketitle
\linespread{1.4}

\textit{This is a pre-copyedited, author-produced PDF of an article accepted for publication in Astronomy \& Astrophysics following peer review. The version of record is available online at: }\url{https://doi.org/10.1051/0004-6361/202346352}.

\begin{abstract}
\emph{Context}. Tidal and rotational deformation of {fluid giant} extra-solar planets may impact {their} transit light curves, making the $k_2$ Love number  observable in the upcoming years. Studying the sensitivity of $k_2$ to mass concentration at depth is thus expected to provide new constraints on the {internal} structure of gaseous extra-solar
   planets.

\emph{Aims}.
   {We investigate the link between the mean 
   {polar} moment of inertia $N$ of a fluid,  stably layered extra-solar planet and its $k_2$ Love number, {aiming at} obtaining analytical relationships valid, at least, for some particular ranges of the model parameters. We also {seek a} general, approximate relationship useful to constrain $N$ once {observations} of $k_2$ will {become} available.}

\emph{Methods}.
   {For two-layer {fluid} extra-solar planets, {we explore the relationship between {$N$ and $k_2$} by analytical methods, for  particular values} of the model parameters. {We also explore} approximate relationships valid over all the possible range of two-layer models. More complex planetary structures are investigated {by the} semi-analytical {propagator technique}.}

\emph{Results}.
   {A unique relationship between $N$ and $k_2$ cannot be established. However, our numerical experiments show that a~\textit{rule of thumb} can be inferred, valid for complex, randomly layered stable planetary structures. The rule robustly defines the upper limit to the values of $N$ for a given $k_2$, and agrees with analytical results for a polytrope of index one
   {and with a realistic non-rotating model of the tidal equilibrium of Jupiter.} }


\end{abstract}


\section{Introduction}


Recent work suggests that the study of transit light curves 
of extra-solar planets may provide information upon their  shape, which is linked to the value of the second degree fluid Love number $k_2$
\citep[see \emph{e.g.,}][]{Carter_2010,correia2014transit,kellermann2018interior,hellard2018measurability,hellard2019retrieval,akinsanmi2019detectability,barros2022detection}. According to~\citet{padovan2018matrix}, estimates of $k_2$ for extra-solar planets may become available in the near future, in view of the expected improvements in the observational facilities and the increasing amount of data. Since for a fluid-like giant planet $k_2$ {is sensitive} to the {density layering}
\citep[\emph{e.g.,}][]{ragozzine2009probing,kramm2011degeneracy,padovan2018matrix}, transit observations {may potentially provide}, in the upcoming years, new constraints {on the internal structure of exoplanets}. These {will have} important implications upon our knowledge of the internal planetary dynamics and the 
formation history~\citep[\emph{e.g.,}][]{kramm2011degeneracy}. 

Using a matrix-propagator approach borrowed from global geodynamics, \citet{padovan2018matrix} have computed numerically the fluid $k_2$ Love number for planetary models of
increasing complexity, ranging from a two-layer 
to {multi-layered} structures. 
Padovan and colleagues have {seen} that the normalised {mean polar} moment of inertia of a planet 
 and $k_2$ show a similar sensitivity to {the mass concentration, \emph{i.e.} they both decrease with increasing mass concentration at depth}, thus supporting 
{the results of}~\citet{kramm2011degeneracy}. The theory 
developed by \citet{padovan2018matrix} is strictly suitable for {close-in}, tidally locked gaseous extra-solar planets, for which the first experimental determinations 
{of $k_2$}  are expected due to their {large size and 
flattening} \citep{hellard2018measurability}. The $N$-$k_2$ 
relationship has never been explored for Earths or super-Earths 
{that include} layers of finite rigidity and {are}  
{less deformable} than gaseous planets \citep{hellard2019retrieval}. 

 {In this work} we delve further into the $N$-$k_2$ relationship for a fluid multi-layered extra-solar planet, with the purpose of refining the implicit approximation of \citet{padovan2018matrix}, namely ${N}\approx {k}_2$. Following these authors, we first adopt a basic two-layer planet, 
and taking advantage of the closed-form expression for $k_2$ first 
published by 
\citet{ragazzo2020theory}, we 
show that an extremely simple power-law (\emph{rule of thumb}) better
captures the relationship between ${N}$ and 
${k}_2$. Second, by { running } a Monte Carlo simulation, we show that for multi-layered models the~\emph{rule of thumb} 
determines an upper limit for $N$ for a given, hypothetically 
observed $k_2$ value. In both cases, the rules obtained are superior to the Radau-Darwin formula  \citep[\emph{e.g.,}][]{cook1980interiors}. 

This paper is organised as follows. In Section~\ref{sec:results}
we recall some basic analytical results regarding the $k_2$ Love number and $N$ for a two-layer fluid planet. 
{In Section~\ref{sec:relationship}} 
{we discuss a possible approximate relationship between $N$ and $k_2$ for a two layer model,}
and test its validity for multi-layered planets through a suite of numerical experiments. {Finally, we draw our conclusions in Section \ref{sec:conclusions}}. 

\section{Analytical results for a fluid two-layer planet}\label{sec:results}

\subsection{$k_2$ Love number}\label{sec:k2}
In the {special case} of a fluid planet, $k_2$ only depends upon the density profile.  The equilibrium equations reduce to a linear, second order differential equation for the perturbed gravitational  potential $\varphi$ that reads:
\begin{equation}\label{eq:varphi}
 {\varphi}^{\prime\prime} + \frac{2}{r} \varphi{^\prime} - \left( \frac{n(n+1)}{r^2} 
+ \frac{4\pi G}{g_0} {\rho}_0^\prime \right) \varphi =0\,,  
\end{equation}
where the prime denotes the derivative with respect to radius $r$, $n$ is the harmonic degree, $g_0(r)$ is 
 gravity acceleration and $\rho_0(r)$ is  {density}~\citep{wu1982viscous}\footnote{{Note that in Eq.~(46a) of ~\cite{wu1982viscous} $\rho_0$ should be substituted by ${\rho}_0^\prime$.}}.
Assuming layers of constant density (\emph{i.e.,} ${\rho}_0^\prime=0$), Eq.~(\ref{eq:varphi}) allows for a closed-form solution 
in terms of powers of $r$. For non-fluid planets that include elastic or visco-elastic layers, a full set of six {spheroidal} equilibrium equations must be solved, since in this case 
  $\varphi$ is coupled with the tide-induced displacements 
\citep[see~\emph{e.g.,}][]{wu1982viscous,melini2022computing}.

Denoting by $r_c$ and $\rho_c$ the radius of the inner layer (the ``core'') and its density, respectively, and by $r_m$ and $\rho_m$ the corresponding quantities for the outer layer (the ``mantle'') {with the aid of the \emph{Mathematica\copyright} \citep{ram2010} 
symbolic manipulator} for $n=2$ we find 
\begin{equation}\label{eq:k2}
\tilde{k}_2 = 2 \frac{    5 + \alpha\, \Big( 5\alpha z^8 + 8 \left(1-\alpha\right)z^5   + 3\alpha    -8 \Big)}
{10 + \alpha \,\Big(  9 z^5 \left(\alpha-1\right)  + 5z^3 \left(5-3\alpha\right) + 6\alpha-16\Big)}\,,
\end{equation}
where $\tilde{k}_2$ is the normalised Love number 
\begin{equation}\label{eq:k2norm}
   \tilde{k}_2 = \dfrac{k_2}{k_{2h}} 
\end{equation}
and
\begin{equation}
k_{2h}=\dfrac{3}{2}   
\end{equation}
is the Love number {for a} homogeneous planet
\citep[see \emph{e.g.,}][]{munk1975rotation}. In (\ref{eq:k2}) 
we have introduced the non-dimensional {core} radius 
\begin{equation}\label{eq:zalfa}
z=\dfrac{r_c}{r_m} \,,
\end{equation}
{with $0\le z \le 1$, and the ratio}   
\begin{equation}\label{eq:alfa-definizione}
    \alpha=\dfrac{\rho_c - \rho_m}{\rho_c}\,.
\end{equation}
\noindent {We note that for a gravitationally stable planet ($\rho_c \ge \rho_m$) we have  
$0\le \alpha \le 1$}. {The value} $\alpha=1$ corresponds to the  {limit case} of a mass-less mantle ($\rho_m=0$), whereas for a homogeneous planet ($\rho_m=\rho_c$), one has $\alpha=0$. 

{Since the planet is fluid and inviscid}, vertical displacement is interpreted as the displacement of equi-potential 
surfaces so that the vertical Love number is $h_2=1+k_2$. As the tangential displacement is undetermined within a perfect fluid, the $l_2$ Love number is undefined.  Further, $k_2'=k_2-h_2$, where $k_2'$ is the loading Love number {for gravitational} potential \citep{molodensky1977relation}. Hence 
$k'_2+1=0$, which manifests a condition of perfect isostatic equilibrium \citep[see \emph{e.g.,}][]{munk1975rotation}. By symbolic manipulation, it is also possible to obtain a general closed-form expression {for} $\tilde k_n$ {at} harmonic degrees $n\ge 2$, which is reported, probably for the first time, in 
Appendix~\ref{sec:appendix-a}. 
It is worth to remark that, although in Eq.~(\ref{eq:k2}) 
$\tilde{k}_2$ is written in terms of  $\alpha$ and 
$z$, {it depends implicitly} upon the four parameters {defining the} model (namely, $r_c$, $r_m$, $\rho_c$ and $\rho_m$). Thus, even assuming that the size of a hypothetical extra-solar planet is known and that we dispose of an observed 
value of $\tilde{k}_2$, it is impossible to determine  the remaining three quantities unambiguously.  

{As far as we know, for the two-layer model, {the} explicit form of $\tilde{k}_2$ has been first published by \citet[][]{ragazzo2020theory}}. In fact, it is easily verified that our {Eq.}~(\ref{eq:k2}) is equivalent to his Eqs.~(2.40) and (2.41), taking into account that {he} has defined $\alpha$ as  $\rho_m/\rho_c$.  
Although \citet{padovan2018matrix} did not provide the  explicit form for $k_2$, we have verified that (\ref{eq:k2}) can be obtained through symbolic manipulation from their analytical propagators, and that it is also consistent, to a very high numerical precision, with the output from the Python codes that they have made available. Furthermore, by symbolic manipulation, we have verified that (\ref{eq:k2}) is also confirmed taking the limit of vanishing frequency {} when the full set of six equilibrium equations for a general visco-elastic layered body  are algebraically solved. A fully numerical computation using the Love numbers calculator \texttt{ALMA}$^3$ of 
\citet{melini2022computing} also confirms (\ref{eq:k2}) to a very high precision. 

As expected, the well-known result $\tilde{k}_2=1$ valid for the Kelvin sphere
\citep{thomson1863xxvii}, is retrieved from Eq.~(\ref{eq:k2})
whenever one of the three limits $\alpha \mapsto 0$, $z\mapsto 0$ 
and $z\mapsto 1$ are taken. The smallest possible value of $k_2$ 
is met in the extreme condition of a point-like mass concentration 
at the planet centre \citep[Roche model, see][]{roche-1873}.
Indeed, with $\rho_m \ll \rho_c$ (hence $\alpha \mapsto 1$) and $z\mapsto 0$, Eq.~(\ref{eq:k2}) gives $k_2 \mapsto 0$, in agreement with  \citet{padovan2018matrix}. 
In Figure~\ref{Fig-k-n}a, the normalised Love number $\tilde{k}_2$ is shown as a function of $\alpha$ and $z$ for the two-layer model, according to Eq.~(\ref{eq:k2}). It is apparent that, for a given $\alpha$ value, the same value of $\tilde{k}_2$ may be obtained for two distinct values of $z$. On the contrary, for a given $z$, knowledge of $\tilde{k}_2$ would determine $\alpha$ unequivocally. However, due to the definition of this parameter {(Eq.~\ref{eq:alfa-definizione})},  
knowledge of $\alpha$ would not suffice  to determine the {layers densities}.

\subsection{Mean polar moment of inertia}

   \begin{figure*}
   \centering
   \includegraphics[width=\columnwidth]{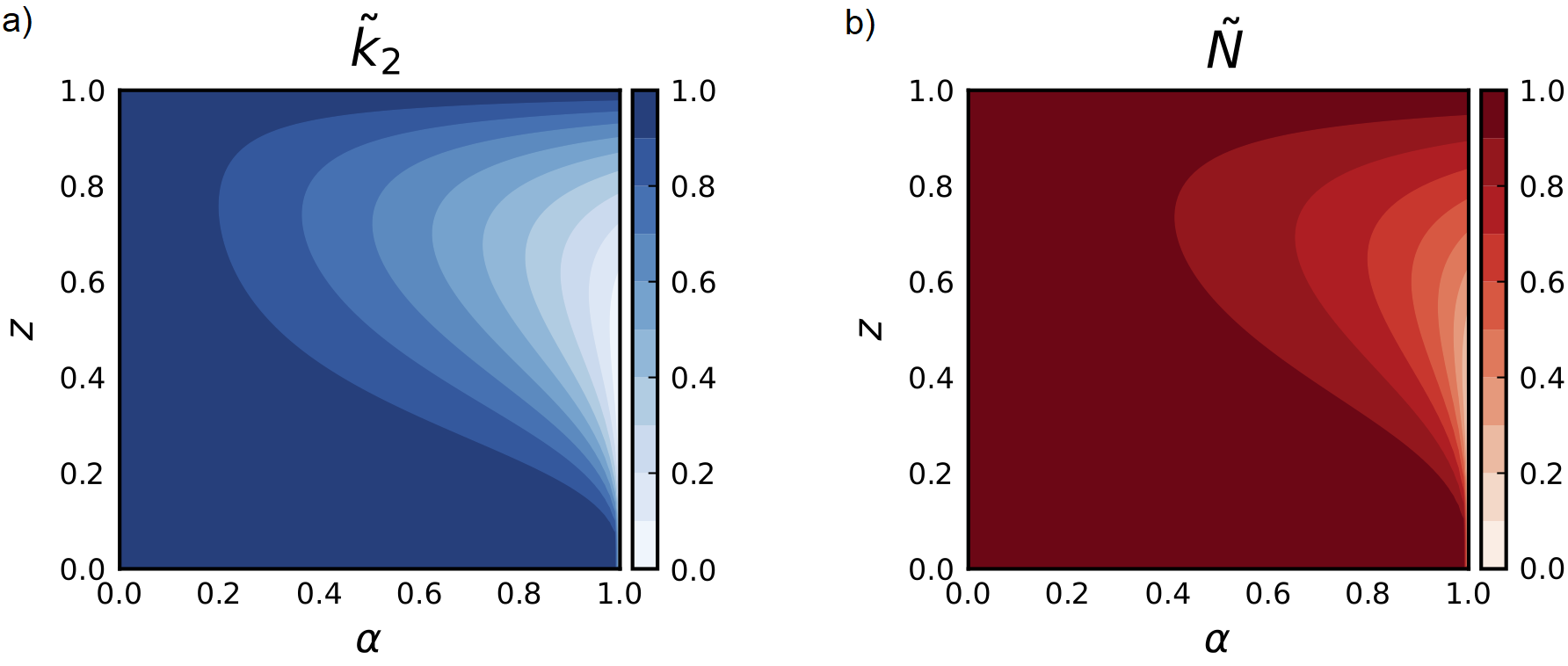}
   \caption{Contour plots showing $\tilde{k}_2$ (a) 
   and $\tilde{N}$ (b) as a function of parameters $\alpha$ and $z$ for a two-layer fluid planet, according to Eqs.~(\ref{eq:k2}) and (\ref{eq:nmoi}), respectively. Since {these variables}
   are normalised {to} the values attained in the case of a homogeneous planet, they both range in the interval ($0,1$).} 
              \label{Fig-k-n}%
    \end{figure*}

Following \citet{ragozzine2009probing}, \citet{hellard2019retrieval}, \citet{kramm2011degeneracy} and \citet{padovan2018matrix}, here we pursue the idea that $k_2$ is a {useful indicator} of the mass concentration at depth inside a planet. 
It is well known that the radial density distribution is {characterised by the normalised polar moment of inertia} 
\begin{equation}
{N}=\frac{C}{MR^2}\,,
\end{equation}
{where $C$ is the 
polar moment of inertia, $M$ is the mass of the body and $R$ is the mean radius} 
{
\citep[see \emph{e.g.,}][]{hubbard1984planetary}.}
The larger is mass concentration {at depth}, the smaller is $N$. Indeed, by its own definition, ${N}$ vanishes in the case of a point-like mass, while for a homogeneous sphere 
\citep[\textit{e.g.},][]{cook1980interiors} it attains the well-known value 
\begin{equation}
    N_h=\dfrac{2}{5}\,.
\end{equation} 

By defining 
\begin{equation}
 {\tilde N} = \dfrac{N}{N_h}\,, 
\end{equation}
for a two-layer planet simple algebra provides 
\begin{equation}\label{eq:nmoi}
{\tilde N} = \frac{ 1 +  \alpha\,(z^5 - 1) }{ 1 +  \alpha\,(z^3 - 1) } \,,
\end{equation}
showing that $\tilde{N}$ and $\tilde{k}_2$ depend upon the same parameters  
$\alpha$ and $z$ but in different combinations ({see}~Eq.~ \ref{eq:k2}), which suggests that establishing the 
$\tilde{N}$-$\tilde{k}_2$ relationship 
{may be not} straightforward. 

We further note that, in analogy with $\tilde{k}_2$, knowledge of $\tilde{N}$ would not allow to invert (\ref{eq:nmoi}) for $\alpha$ and $z$ unequivocally (hence for the four model parameters $r_c, r_m, \rho_c$ and $\rho_m$), unless further constraints are invoked. Based upon Eq.~(\ref{eq:nmoi}), in Figure~\ref{Fig-k-n}b  the ratio $\tilde{N}$ is shown as a function of parameters $\alpha$ and $z$. As noted for $\tilde{k}_2$ in
Figure~\ref{Fig-k-n}a, for a given value of $\alpha$ the same $\tilde{N}$ can be obtained for two distinct values of $z$, while for a given $z$ knowledge of $\tilde{N}$ would determine 
$\alpha$ unequivocally.

\section{Relationship between the moment of inertia and the $k_2$ fluid Love number}\label{sec:relationship}

\subsection{{Two-layer models}}\label{sub:approx}

 \citet{padovan2018matrix} have established a method for the evaluation of $\tilde{k}_2$
for a general fluid planet, based upon the propagator technique often employed in geodynamics \citep[\emph{e.g.,}][]{wu1982viscous}.  
Furthermore, following the work of 
\citet[][]{kramm2011degeneracy}, they have shown that for a planet with two constant density fluid layers, $\tilde{N}$ and $\tilde{k}_2$ are directly correlated, {both decreasing with} increasing mass concentration at depth. However, \citeauthor{padovan2018matrix} did not 
propose explicitly a general relationship between these two 
{quantities},
which they enlightened  for particular planetary models characterised by a specific mass, size and density  
(see their Figure $3$). 

On one hand, by comparing Figure~\ref{Fig-k-n}a with \ref{Fig-k-n}b it is apparent that, for our two-layer model, 
functions $\tilde{k}_2$ and $\tilde{N}$ have broadly similar shapes in the $(\alpha,z)$ plane, 
immediately suggesting a straightforward {linear} relationship $\tilde{N} \simeq \tilde{k}_2$. 
Such relationship has been implicitly proposed by \citet{padovan2018matrix}  and would be exact for a uniform sphere. 
On the other hand,
if we limit ourselves to an inspection of {the analytical} expressions  ($\ref{eq:k2}$) and (\ref{eq:nmoi}), {it is not easy to guess whether an exact $\tilde{N}$-$\tilde{k}_2$ relationship may exist in analytical form.} 
\emph{A priori,} 
 for a non-homogeneous planet such relationship 
{might} be non-univalent, with more $\tilde{N}$ values {corresponding} to a given $\tilde{k}_2$ and \emph{vice-versa}. 
    
After some symbolic manipulations, we have verified that solving Eq.~(\ref{eq:nmoi}) for $\alpha$ and substituting into (\ref{eq:k2}) would 
{not provide insightful results}.
This suggests that an exact relationship $\tilde{N}=\tilde{N}(\tilde{k}_2)$
not involving $\alpha$ and $z$ explicitly and valid for all values of these parameters can be almost certainly ruled out. 
Nevertheless, simple relationships of partial validity could exist in some 
limiting cases where $\alpha$ or $z$ take special values.
{For example, it is easy to show that for small core bodies 
($z \mapsto 0$), $\tilde{N} \simeq 1+ (2/5)( \tilde{k}_2 - 1)$,
which holds {for all values of $\alpha$} and still implies that mass concentration {at depth}
increases for decreasing $\tilde{k}_2$.}
Along the same lines, we note that for $\alpha \mapsto 1$, corresponding to case of a dense ``core'' surrounded by {a ``light mantle''}, Eq.~(\ref{eq:nmoi}) {gives} 
$\tilde{N} \simeq z^2$   
and since from Eq.~(\ref{eq:k2}) 
$\tilde{k}_2 \simeq z^5$,  
by eliminating $z$ we obtain {an} appealingly simple {approximate} power-law relationship
$\tilde{N} \simeq {\tilde{k}_2}^{0.4}$.
{We note that 
this last relationship is actually an exact result for a homogeneous sphere surrounded by an hypothetical zero-density mantle, and can be obtained analytically by re-scaling the results for a Maclaurin spheroid \citep{hubbard2013maclaurin} of radius $a$  
to the outer radius $r>a$ of the mass-less envelope
(Hubbard, 2023, personal communication).} 


The approximate $\tilde{N}$-$\tilde{k}_2$ relationships {discussed above} are only valid for specific ranges of $\alpha$ and $z$. 
Certainly, 
{a straightforward linear relationship}
captures the broad similitude of the diagrams in Figures~\ref{Fig-k-n}a and \ref{Fig-k-n}b, but {it may represent} a too simplistic solution. 
{Here, we seek a more} general \emph{rule of thumb} (or ROT) providing, within a certain level of approximation, a {relationship} {between $\tilde{N}$ and $\tilde{k}_2$ over} all the points of the $(\alpha,z)$ plane. To quantify the error associated to a given ROT (say, $\tilde{N}_{ROT}(\tilde{k}_2)$), we introduce the non-dimensional root mean square 
\begin{equation}\label{eq:RMS}
\mathrm{RMS} = \sqrt{\int_0^1 \!\! \int_0^1 
\left[ \tilde{N}  - \tilde{N}_{ROT}(\tilde{k}_2) \right]^2 
 d\alpha \, dz} \,,
\end{equation}
where the double integral {is} evaluated numerically by standard methods. 

\begin{figure}
   \centering
    \includegraphics[width=0.8\columnwidth]{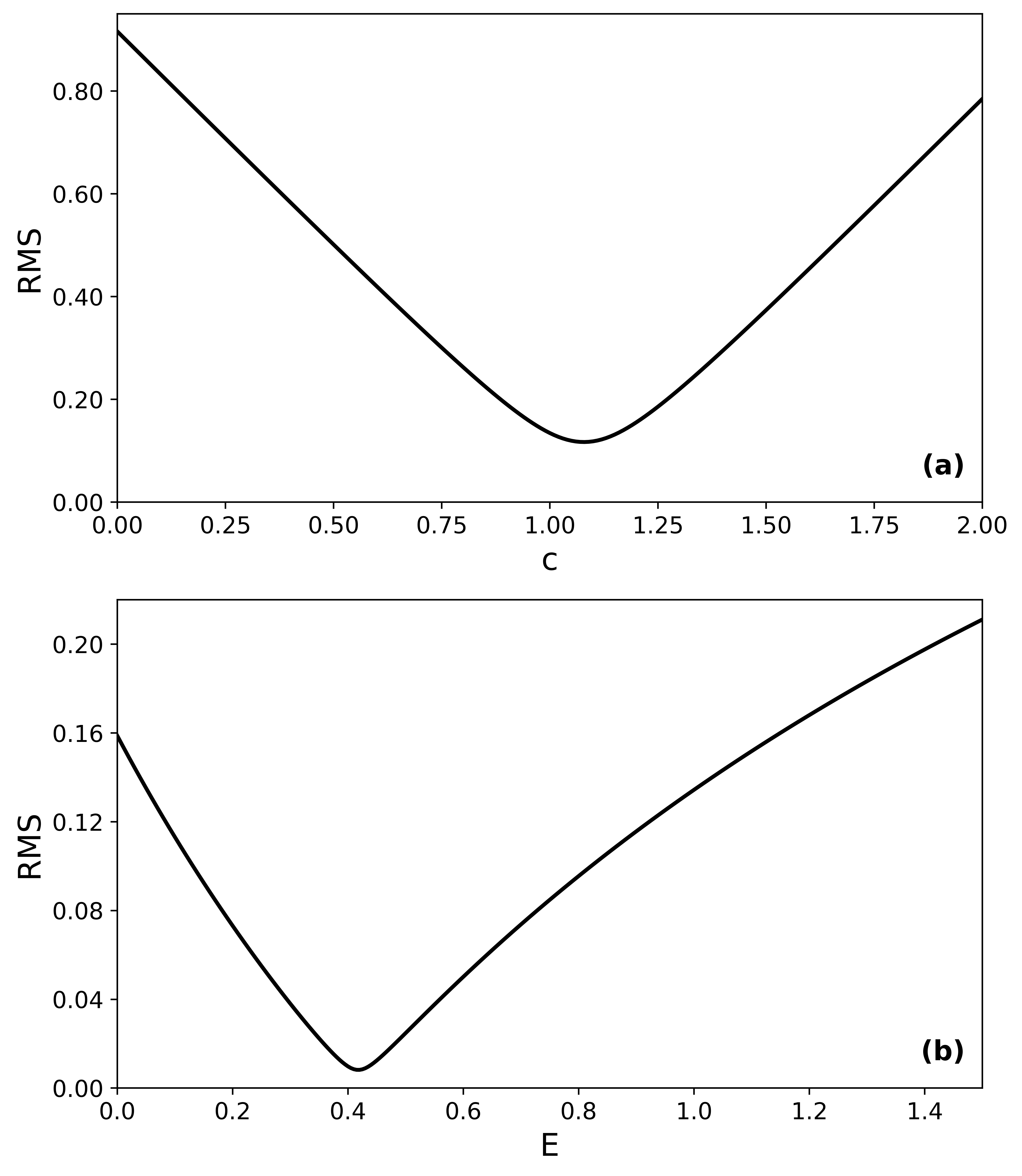}
   \caption{Non-dimensional RMS, evaluated according to 
   Eq.~(\ref{eq:RMS}), for a linear ROT $\tilde{N}\approx c\tilde{k}_2$ (frame a) and for a power law ROT $\tilde{N}\approx \tilde{k}_2^E$ (b), as a function of the parameters $c$ and $E$, respectively. Integrals in Eq.~(\ref{eq:RMS}) have been evaluated numerically {by} the~\texttt{dblquad} function included in the SciPy library \citep{2020SciPy-NMeth}.} 
              \label{Fig-misfit-rms}%
    \end{figure}

{First, we assume a direct proportionality}
\begin{equation}\label{eq:rot_linear}
\tilde{N} = c\,\tilde{k}_2\,,
\end{equation}
where $c>0$ is a constant. 
Figure~\ref{Fig-misfit-rms}a 
shows, as a function of $c$, the RMS obtained with $\tilde{N}_{ROT}=c\,\tilde{k}_2$. The minimum RMS (close to 0.1168) is obtained for $c\approx 1.08$, suggesting that the {approximation 
{$\tilde{N}\simeq \tilde{k}_2$} 
proposed} by \cite{padovan2018matrix} and 
corresponding to $c=1$, is indeed close to the best possible linear ROT.

Next, we consider a power-law relationship 
\begin{equation}\label{eq:thumbE}
\tilde{N} = {\tilde{k}_2}^E \,,
\end{equation} 
where $E>0$ is an adjustable exponent. 
In Figure~\ref{Fig-misfit-rms}b  we show, as a function of $E$, the RMS corresponding to $\tilde{N}_{ROT}=\tilde{k}_2^E$.
It is apparent that the RMS is minimised for 
{an exponent} $E \approx 0.42$, close to the value of $0.4$ {found analytically for a zero-density mantle}.
The corresponding minimum RMS value is $\approx 0.0082$. 
{These findings} suggest that the relationship  
\begin{equation}\label{eq:thumb04}
 \tilde{N} \approx {\tilde{k}_2}^{\,0.4}
\end{equation} 
{represents} a simple and valid ROT expressing the 
link between $\tilde{N}$ and $\tilde{k}_2$ for a two-layer, fluid, {stably layered} planet characterized by {arbitrary} parameters $\alpha$ and $z$. 

\subsection{{Arbitrarily layered models}}

{Up to now, we  have limited our attention to} four-parameters models composed by two distinct fluid layers. To fully assess the validity of the ROT {(\ref{eq:thumb04})},
it is {important} to consider {the case} of a planetary structure consisting of an arbitrary number 
$L$ of homogeneous layers. 

{Due to the model complexity}, 
in this general case {an analytical expressions for $\tilde{k}_2$ is not available}; however, it is possible to evaluate $\tilde{k}_2$ numerically, for instance following the propagator method outlined by~\cite{padovan2018matrix} or~ employing numerical~ Love~ ~numbers ~calculators~ like~\texttt{ALMA} \citep{melini2022computing}. Conversely, an analytical expression for the normalised moment of inertia $\tilde{N}$ is easily obtained also in the general case of an $L$-layer planet, {and it reads}
\begin{equation}\label{eq:nmoi2}
    \tilde{N} = \frac{ \displaystyle
    \sum_{i=1}^{L} (1-\alpha_i) \left( {z_i^5-z_{i-1}^5} \right)}
    {
    \displaystyle\sum_{i=1}^L (1-\alpha_i) 
    \left(z_i^3-z_{i-1}^3\right) }\,,
\end{equation}
where $z_i=r_i/r_m$ {is the normalised radius of the outer boundary of the $i$-th layer} 
{($z_0 \equiv 0$)} and 
\begin{equation}
\alpha_i = \frac{\rho_1-\rho_i}{\rho_1}\,,
\end{equation}
{where $\rho_i$ is} the density of the $i$-th layer. By definition, $z_1 \le \ldots \le z_L=1$, {while gravitational stability imposes} $\rho_1 \le \ldots \le \rho_L$ {so that} $1\ge \alpha_L \ge \ldots \ge \alpha_1= 0$. It is easily shown that, for $L=2$, Eq.~(\ref{eq:nmoi2}) reduces to (\ref{eq:nmoi}) with $\alpha\equiv\alpha_2$ and $z\equiv z_1$.

\begin{figure}
   \centering
\includegraphics[width=0.8\columnwidth]{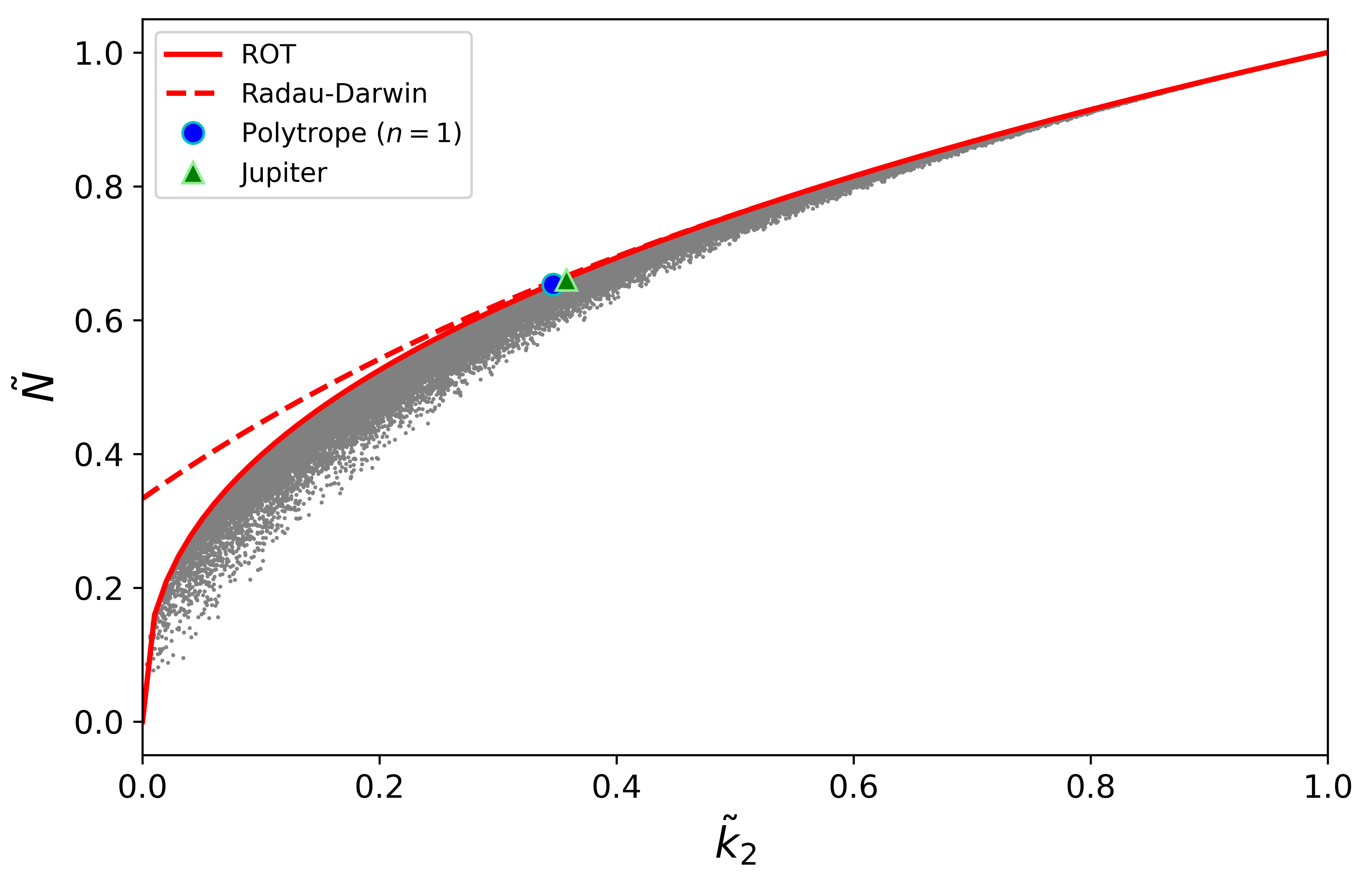}
      \caption{Fluid Love number $\tilde{k}_2$ and normalised moment of inertia $\tilde{N}$ for a random ensemble of $5\times 10^5$ models with a number of layers {$2\le L \le 10$}.  {The solid line {shows} the ROT $\tilde{N}=\tilde{k}_2^{0.4}$}. {The dashed one represents the Radau-Darwin (RD)  formula~\citep[\emph{e.g.,}][]{cook1980interiors,padovan2018matrix,ragazzo2020theory}}. ~{The RD formula is exact for a homogeneous body  but it constitutes an approximation for layered planets~\citep{kramm2011degeneracy,padovan2018matrix}.}
{The ROT and the RD formula match for $\tilde{k}_2 \gtrsim 0.3$;
for smaller values, our ROT represents a more rigorous upper limit to $\tilde{N}$.}
      The blue dot corresponds to values of $\tilde{k}_2$ and $\tilde{N}$ for a polytrope of index one, 
      {while the green triangle corresponds to results by \citet{wahl2020jupiter} for the equilibrium tidal response of Jupiter}. 
              }
         \label{fig:random_models}
   \end{figure}

{To test} whether the ROT (\ref{eq:thumb04}) can be of practical use also for general planetary structures, we have generated an ensemble of $5\times10^5$ models with a number of layers variable between $L=2$ and $L=10$, all characterised by a gravitationally stable density profile. For each of the planetary structures so obtained, we have computed $\tilde N$ according to Eq.~(\ref{eq:nmoi2}) and $\tilde{k}_2$ with the numerical codes made available by~\citet{padovan2018matrix}. The corresponding values of $\tilde{N}$ and $\tilde{k}_2$ are shown in
Figure~\ref{fig:random_models} as grey dots. 

For a given, hypothetically observed $\tilde k_2$
value, the corresponding value of $\tilde N$ is clearly not unique. Rather, $\tilde N$ ranges within an interval, 
defined by the cloud of points, whose width represents the uncertainty 
associated to the degree of mass concentration at depth. It is apparent that the maximum relative uncertainty on $\tilde N$ (up to $\sim 50\%$)
occurs for $\tilde k_2$ values $\lesssim 0.2$ and that, for $\tilde k_2$ exceeding $\approx 0.5$, the $\tilde N$ value 
is rather well constrained (to within $\approx 10\%$). Of course, this does not imply that the density profile of the planet is actually constrained, since 
Eq.~(\ref{eq:nmoi2}) cannot be inverted for $\alpha_i$ and $z_i$
unequivocally without introducing further assumptions.
The solid red line in Figure \ref{fig:random_models} represents the ROT (\ref{eq:thumb04}),
obtained in the context of the two-layer model in Section
~\ref{sec:k2}. It is apparent that the ROT remains valid also in the general case of a $L$-layer planetary model and, for $\tilde{k}_2 \gtrsim 0.5$, it provides a good estimate of $\tilde{N}$ once $\tilde k_2$ is known. For smaller values of $\tilde{k}_2$, the ROT represents an upper bound to the normalised moment {of inertia:}
\begin{equation}\label{eq:upperlimit}
\tilde{N} \lesssim \tilde{k}_2^{0.4}\,.
\end{equation}

{In the context of planetary structure modelling, the polytrope ~of unit index \citep{chandrasekhar1933} has a particular relevance. This ~simplified ~model ~resembles the interior barotrope of a hydrogen-rich planet in the Jovian mass range and, by virtue of its linear relationship between mass density and gravitational potential, it allows for the derivation of exact results useful for calibrating numerical solutions. \citet{Hubbard_1975} obtained analytical expressions of the moment of inertia and of the $k_2$ fluid Love numbers for a polytrope of index one, which are marked by a blue dot in Figure \ref{fig:random_models}. More recently, \citet{wahl2020jupiter} modelled the equilibrium tidal response of Jupiter through the concentric Maclaurin spheroid method; their results in the non-rotating limit are marked by a green triangle in Figure \ref{fig:random_models}. It is evident that the ROT turns out to be in excellent agreement with these two particular cases.}

\section{Conclusions}\label{sec:conclusions}
In this work, we have re-explored the relationship between the Love Number $\tilde k_2$ of a fluid extra-solar planet and its mean {polar} moment of inertia $\tilde N$. Such a relationship would allow, in principle, an indirect inference  of constraints on the internal mass distribution on the basis of an observational determination of $\tilde k_2$. {However, we remark that for a quantitative application of our results to real exoplanets, rotational effects and  nonlinear responses to rotational and tidal terms should be also considered \citep[see, \emph{e.g.}][]{wahl2017macalurin,wahl2020jupiter}}. 

{Our conclusions are twofold.} 

\emph{i).} For a hypothetical planet consisting of two homogeneous fluid layers, using the exact propagators method, we have confirmed that a relatively smooth analytical expressions of $\tilde k_2$
can be found. However, this expression does not allow to establish a {unique} analytical relationship between $\tilde k_2$ and $\tilde N$, except for some particular ranges of the model parameters. By investigating some approximate relationships, for the first time we have determined the \textit{rule of thumb} $\tilde N \approx \tilde k_2^{0.4}$, which is providing a good estimate of $\tilde N$ as a function of $\tilde k_2$ over the whole range of possible two-layer models. 

\emph{ii).} {By a Monte Carlo approach}, we have explored the validity of our ROT in the general case of gravitationally stable planetary models with an arbitrarily large number of homogeneous layers. We have found that {the} ROT provides an \emph{upper limit} to the possible range of mean moment of inertia corresponding to a given value of $\tilde k_2$, {and the distribution of downward departures from ROT {increases} as $\tilde{k}_2 \mapsto 0$}. In addition, the ROT is in good agreement with analytical results for a fluid polytrope body of unit index {and with a realistic non-rotating model of the tidal deformation of Jupiter}.  Remarkably, our simulations show that especially for small values of $k_2$, the ROT is more accurate than {the celebrated {RD} formula}.


\section*{Acknowledgements}
      {We thank Bill Hubbard for his insightful review that greatly helped to improve the original manuscript}. We are {indebted} to Roberto Casadio for {discussion} and to {Leonardo Testi and} Andrea Cimatti for encouragement. {We also thank Nicola Tosi for advice}.  {AC and GS are} supported by a {``RFO''} DIFA grant.

%
\bibliographystyle{aa} 
\bibliography{scibib-GIORGIO-new.bib} 
%
\begin{appendix}\label{app:kn} 
\section{Analytical expression of $k_n$ for a two-layer fluid model}\label{sec:appendix-a}

Here we give an analytical expression for the tidal Love number of degree $n\ge 2$ for a fluid two-layer extra-solar planet. {Consistent with 
(\ref{eq:k2norm})}, we introduce a normalised Love number:
\begin{equation}
    \tilde{k}_n = \dfrac{k_n}{k_{nh}}\,,
\end{equation}
where
\begin{equation}
k_{nh}=\dfrac{3}{2(n-1)}   
\end{equation}
is the Love number for a homogeneous planet. With the aid of the \emph{Mathematica\copyright} \citep{ram2010} symbolic manipulator 
we obtain the exact solution 
\begin{equation}\label{eq:kN}
\tilde{k}_n = 
2(n-1)
   \frac{
   \alpha ^2 (2 n+1) z^{2(n+2)}
   +2 \alpha (1-\alpha) (n+2) z^{2n+1}
   +(1-\alpha) (2n+1-3 \alpha)
   }{
   9 (\alpha -1) \alpha  z^{2n+1}
   +\alpha  (2 n+1) (2n+1-3\alpha) z^3
   +2 (1-\alpha)
   (n-1) (2n+1-3 \alpha) 
   }\,.
\end{equation}
It is easily verified that, for $n=2$, (\ref{eq:kN}) reduces to (\ref{eq:k2}), and that for $\alpha\mapsto 0$, $z \mapsto 0$ and $z \mapsto 1$ the homogeneous limit $\tilde{k}_n = 1$ is obtained.

\end{appendix}

\end{document}